# MILLISECOND X-RAY PULSES FROM CYGNUS X-1

J. F. Dolan

Department of Astronomy

San Diego State University

San Diego, CA  92182-1221

tejfd@sciences.sdsu.edu

# ABSTRACT

X-ray pulses with millisecond-long FWHM gave been detected in RXTE (Rossi X-Ray Timing Explorer) observations of Cyg X-1. Their identity as short-timescale variations in the X-ray luminosity of the source, and not stochastic variability in the X-ray flux, is established by their simultaneous occurrence and similar pulse structure in two independent energy bandpasses. The light-time distance corresponding to the timescale of their FWHM indicates that they originate in the inner region of the accretion disk around the system's black hole component. The fluence in the pulses can equal or exceed the fluence of the system's average continuous flux over the duration of the pulses' FWHM in several different bandpasses between 1 and 73 keV. Millisecond pulses are detected during both high and low luminosity states of Cyg X-1, and during transitions between luminosity states.

1. INTRODUCTION

Sunyaev (1973) pointed out that turbulent convection or magnetic reconnection should lead to hot spots – self-luminous clumps of emitting material – in a quiescent accretion disk surrounding a black hole. A quiescent disk, radiating solely by thermal processes, occurs when the mechanisms that transfer angular momentum outward in the disk have low efficiency, parameterized as $\alpha < 1$, where $\alpha = [v_t/v_s] + [H^2/4\pi\rho v_s^2]$, and $v_s$ is the speed of sound, $v_t$ is the turbulent velocity, H is the magnetic field strength, and $\rho$ is the density in the disk. The luminosity of theses flare patches may equal or exceed the average luminosity of the entire disk in some energy bandpasses.

If the lifetime of a flare patch exceeds its orbital period around the black hole, then Doppler effects and occultation by the black hole will modulate the flux from the flare patch received at the Earth with a timescale characteristic of this orbital period. A typical maximum to minimum luminosity ratio for the flux is calculated to be 16 (Sunyaev 1973; Cunningham & Bardeen 1972), so the emitted flux detected at Earth will appear to be a pulse of radiation. Shakura & Sunyaev (1973) estimate the characteristic timescale of these fluctuations as

$$\Delta\tau = (10^{-4}/\alpha)\, m\, (r/3)^{3/2}\ \text{s} \qquad [1]$$

where m is the mass of the black hole in solar masses and r is the distance of the flare patch from the black hole in units of the Schwarzschild radius $R_S = 2GM/c^2$. The innermost stable orbit around a Schwarzschild (non-rotating) black hole has r = 3, so for m = 10 a characteristic timescale > 1 ms is expected for the pulses. Stable orbits do exist for r < 3 for a Kerr black hole with angular momentum (in geometric units) a > 0. For a approaching 1, its maximum value in geometric units, the characteristic timescale is 7 – 10 times shorter than that given in Eq. [1]. i.e., $\Delta\tau \sim 1$ ms (Sunyaev 1973; Novikov & Thorne 1973).

Rothschild et al. (1974, 1977) detected variable X-ray luminosity from Cyg X-1 that they described as bursts of radiation with a millisecond timescale. Detections occurred in a 2–35 keV bandpass during two different rocket flights with 320 µs time resolution in the first flight and 160 µs time resolution in the second. Both observations were made during the 2-10 keV low flux state of Cyg X-1. (The 2-10 keV low flux state is the hard spectral state of the source. It is also its low luminosity state [Dolan et al. 1979].) The average intensity profile of the bursts with this time resolution was consistent with a rectangular ("top-hat") pulse.

The identification of the variability detected by Rothschild et al. in low count-rate data as ms-timescale pulses was questioned by Press & Schachter (1974), who pointed out that stochastic variability in the flux from the system could mimic short-timescale flares if it occurred on top of small, slow variations in the luminosity of the source. Weiskopf & Sutherland (1978) also questioned the identification of the variability as ms pulses because of the presence of shot-noise variability in the X-ray flux of Cyg X-1. Priedhorsky et al (1979) detected no ms bursts in an 84 s long observations of Cyg X-1. These authors also found shot noise to be an acceptable model of its variability.

To further complicate the situation, Giles (1981) reported ms-timescale bursts during a 6.4 s rocket observation of Cyg X-1, but cautioned that the evidence for ms-timescale bursts was not definitive because of the statistical limitations imposed by the small number of counts in each burst. Giles also found his data to be consistent with a shot-noise model of variability. Meekins et al. (1984) also reported variability on a timescale of ~ 3 ms during a 9 minute observation of Cyg X-1 with HEAO-1, modeling it as the superposition of uncorrelated 3 ms and 300 μs shots. But Chaput et al. (2000) found no ms bursts from Cyg X-1 in 1.3 hours of 1-25 keV HEAO-1 and RXTE observations. The latter authors concluded that the Meekins et al. results were spurious, and attributed them to instrumental effects. Uttley & McHardy (2001) then showed that the standard shot-noise models do not apply to the variable flux from Cyg X-1 because time series data from the source is not stationary on short timescales. (The rms variability scales linearly with the flux.)

More recently, Gierlinski & Zdziarski (2003) reported 13 strong ms-timescale flares in over 600 hours of 2-60 keV observations of Cyg X-1 with RXTE. The flares occurred in both the hard and soft spectral states of the source. We report her that ms-timescale pulses are a common occurrence in the 1-15 keV flux from Cyg X-1 observed with RXTE. The pulses occur during both high and low luminosity states, and during the transitions between these states.

## 2. OBSERVATIONS AND DATA ANALYSIS

We analyzed 7.95 hr of archival RXTE observations of Cyg X-1 for millisecond-timescale pulses. The observations were obtained during 1996 February, May, June, August, and December; 1997 September; and 2000 December. During 3.69 hr of this time, Cyg X-1 was in the 2-6 keV high state, which is the soft spectral state / high luminosity state of the source. 1.77 hr of these high-state observations were obtained 2 d before a transition from the high state to the low state, which is the hard spectral state / low luminosity state of the source. 2.69 hr of observations were obtained during failed transitions between the low state and the high state, including 1.02 hr during a drop back from the high state to the low state.

All data were obtained using the Proportional Counter Assembly (PCA) detectors with a time resolution of either 122 μs or 244 μs. The data were binned in two separate energy bandpasses, each of which contained an approximately equal number of photons. The two bandpasses were independent, in the sense that each detected photon appeared in only one of the bandpasses.

We define a pulse as a statistically significant intensity variation having a full width at half-maximum (FWHM) duration shorter than a pre-selected value. The confidence level of the identification of a positive-going variation in counting rate as a pulse is then determined by a statistical estimate of the frequency of occurrence of stochastic variations in the detector's counting rate as large as the pulse or larger. If the number of candidate pulses detected significantly exceeds the expectation value for the number of variations that large or larger above the mean flux given by Poisson statistics, then intensity variations caused by pulses are present in the data. (Note that a frequency of occurrence analysis can never positively identify any individual event as a pulse because, given a long enough observing time, stochastic variability can mimic any characteristic of a photon pulse. A frequency of occurrence analysis can show how likely or unlikely it is that an observed variation is a stochastic variation.)

Because the X-ray flux from Cyg X-1 exhibits stochastic variability over timescales shorter than 1 s, the mean flux to which the significance level of a variation must be referred is the local mean flux of the source in a time interval that includes the variation. The two intervals from which the mean counting rate was estimated included the underlying counts in the wings of any pulse, but excluded the peak count rate channels. If the variation is stochastic and not a pulse, then excluding the peak count rate region from the calculation of the local mean biases the mean to lower values. We therefore adopted a mean flux level 10% above that calculated to attempt to account for this bias. In any case, because of the small probabilities involved, this procedure should discriminate between a stochastic variation and a pulse.

The maximum S/N in a detector's counting rate produced by a pulse with a given fluence will occur when the sample interval of the data equals the FWHM of the pulse and the center of a sample interval is coincident with the maximum of the pulse (Dolan 2001). If the sample interval is longer than the FWHM of the pulse, the number of counts from the continuous flux will increase in proportion to the sample interval while the number of counts in the pulse will remain constant, lowering the S/N of the pulse. If the sample interval is shorter than the FWHM of the pulse, or if the center of the pulse is not the center of the sample interval, then the counts from the pulse will be divided between two or more sample intervals, each of which has, on average, the same number of counts from the continuous flux, also lowering the maximum S/N attained by the pulse in any one sample interval. Searching for ms-timescale pulses with only one sample time introduces a selection effect into the detection process that discriminates against pulses with a FWHM duration different from the time resolution of the observations.

We searched for pulses with FWHM between ~ 0.2 and ~ 10 ms. We therefore rebinned the original data into additional data sets modulo 2, 3, 5, 10, 20, and 40 times the original time resolution (0.122 or 0.244 ms) to improve the detectability of ms-timescale pulses. When we sum the counts in n successive sample intervals of the original data set to produce the counts in one rebinned sample interval, there are n different data sets we can produce by starting the furst sum with sample number 0, 1, 2, . . . n-1 of the original data set. We call the sample number with which we start the first sum the phase of the rebinned data set. Estimates of the statistical significance of any detected pulses must than take into account a factor representing the number of ways the original data set was rebinned.

Additional criteria can now be used to discriminate between pulses and stochastic variations:

- a pulse should exist at the same time in two independent energy bandpasses; stochastic variability should be random in independent energy bandpasses;

- a pulse should exist at multiple phases of a rebinned data set with a given sample time. Stochastic variability displays random structure at different phases of a rebinned data set;

- a pulse should exhibit a characteristic rise to and fall from maximum when examined with a sample time smaller than the pulse's FWHM. Stochastic variability exhibits fluctuations between the counts in adjacent sample intervals that are random in direction.

The requirement that any variability meet these additional criteria in order to be identified as a pulse increases the confidence level of the identification of ms-timescale pulses beyond that resulting from the statistical significance criterion alone.

The FWHM duration of the event, in sample intervals, was estimated as the time at the center of the sample interval on the descending slope of the pulse containing half the counts of the peak sample interval (after subtracting the counts per sample in the continuous flux from each sample) minus the time at the center of the corresponding sample interval on the ascending slope. The uncertainty of the time of half-maximum is one-half a sample interval, so the uncertainty of the FWHM is $\sqrt{2}/2$ times the width of a sample interval. FWHM are calculated using the time resolution of each binning and phase at which a pulse is detected. We use the FWHM with the lowest associated uncertainty as the FWHM of the pulse. The FWHM used in calculating the probability of a stochastic variation producing the fluence of the event is the FWHM determined with the time resolution and phase of the data set in which the probability is calculated.

## 3. RESULTS

Millisecond-timescale X-ray pulses exist in every 100 s duration data set we examined. We list in Table 1 the properties of several such pulses, selected to illustrate the range of parameters they possess.

## Table 1

## Properties of Selected X-Ray Pulses from Cyg X-1

| Event | Luminosity State | Bandpass (keV) | FWHM (ms) | Spectral Hardness Ratio Pulse | Bkg |
|---|---|---|---|---|---|
| 960212 | Low | 1 - 73 | 24 ± 4 | 1.9 ± 0.3 | 2.1 |
| 961217 | Low | 1 – 13 | 1.2 ± 0.6 | 1.4 ± 0.8 | 0.6 |
| 990925A B | Failed low to high transition | 2 - 15 | 10 ± 2<br>10 ± 2 | 1.9 ± 0.3<br>2.0 ± 0.4 | 1.2 |
| 990925C D | " | 2 – 15 | 15 ± 4<br>15 ± 4 | 0.6 ± 0.2<br>0.35 ± 0.25 | 1.2 |
| 960811 | High | 1 – 13 | 1.2 ± 0.9 | 1.8 ± 0.4 | 2.3 |
| 960522 | High | 1 – 13 | 7.3 ± 0.9 | 5.1 ± 0.8 | 2.8 |

### 3.1. Event 960212

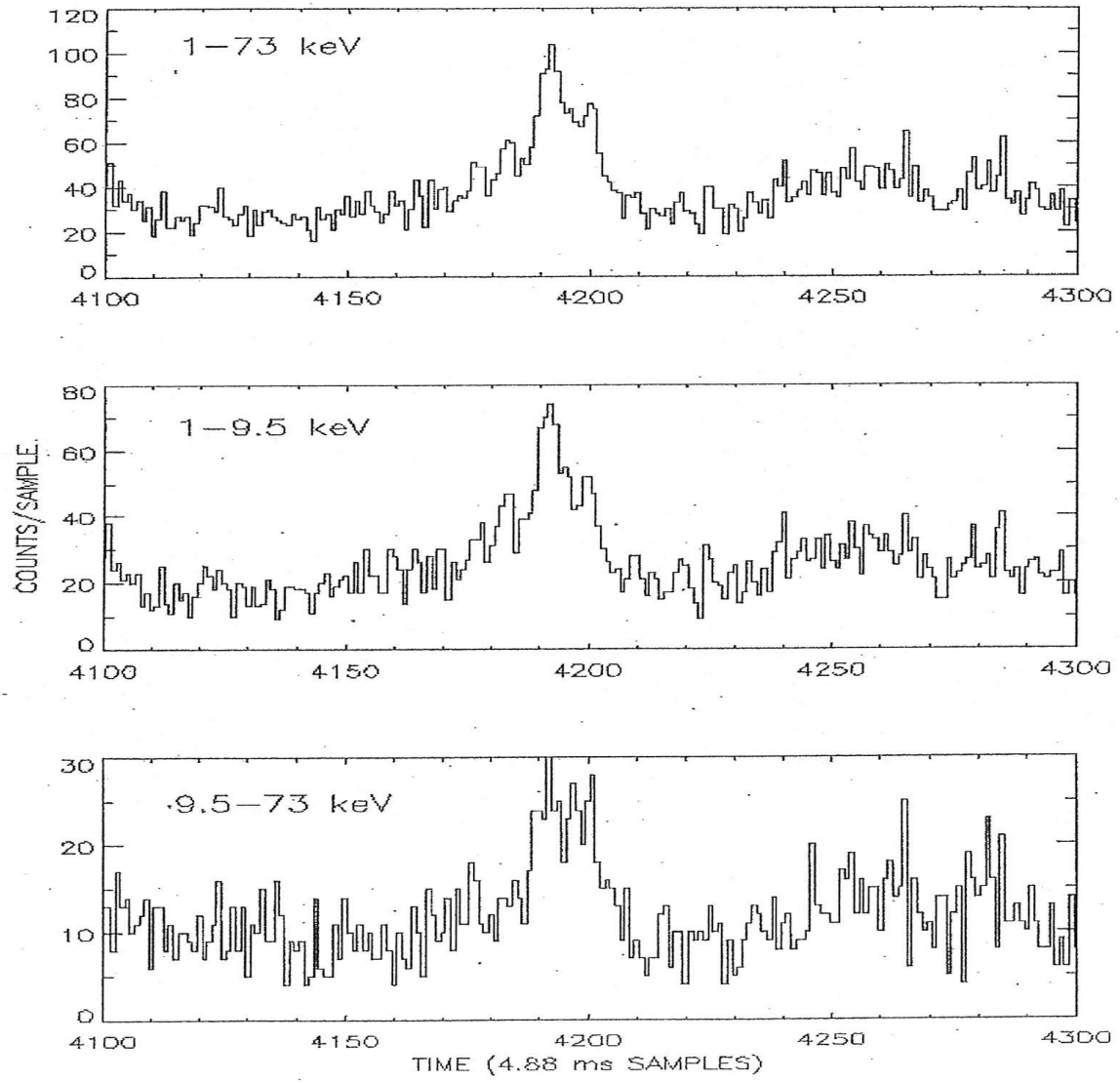

Fig. 1. Event 960212. The PCA counting rate at 4.88 ms sample time. The time is given as the number of sample intervals after the start of the data set. The event peaks in sample interval 4192. *above*: 1-73 keV bandpass; *middle*: 1-9.5 keV; *below*: 9.5-73 keV, all phase 1.

The event occurred on 1996 February 12 at 09:39:20 UT, when Cyg X-1 was in the low state. The counting rate in a 1-73 keV bandpass is shown in Fig. 1, binned at 4.88 ms sample time. The bin numbers on the x –axis are the number of 4.88 ms time intervals since the start of the data set. The event we identify as a pulse peaks at 105 counts in sample interval 4192. We assign a local continuous flux level of 40 counts per sample interval at the time of the event. The Poisson probability of observing a sample interval with 105 counts or more in a data set with a mean of 40 counts per sample is $10^{-23}$. There are 20,480 sample intervals in the 100 s long data set; the expectation value for the number of samples with a count of 105 (or larger) is $1.5 \times 10^{-19}$. If we treat the binning at separate phases as independent trials and multiply by 40, the number of examined phases at 4.88 ms sample time, the probability that the event is a stochastic variation is $10^{-17}$. (We emphasize, however, that the 40 data sets at different phases are not independent. The event appears at a similar level and with a similar shape at all 40 phases. Including this factor overestimates the probability of the event being a stochastic variation.) Multiplying by 288, the number of separate 100 s duration observations of Cyg X-1 we examined, and again by 7, the number of different sample times examined for each 100 s observation (which again are not independent trials), the probability of this event occurring stochastically in any of the data we examined is $< 10^{-14}$.

Additional evidence exists for the identification of this event as a burst of radiation.

- The FWHM of event 960212, given a continuous flux level of 40 counts per sample interval, is 5 sample intervals. The counting rate of the event in these 5 sample intervals, which we define as the fluence of the pulse, is 454 counts. The counting rate expected in the same 5 intervals from the continuous flux is 200 counts. The Poisson probability of obtaining a sample interval with 454 counts in a distribution with an expectation value of 200 counts is $10^{-73}$.

- The event occurs at the same time and with a similar shape in both the 1-9.5 keV bandpass and the 9.5-73 keV bandpass (Fig. 1).

- The event shows a smooth rise to and fall from maximum when examined at faster time resolution.

This event meets all criteria for a ms-timescale burst of radiation from the source.

The fluence of the pulse over its FWHM in the 1-73 keV bandpass is 454 – 200 = 254 counts, equaling the continuous flux from the source over the duration of the pulses' FWHM. Assigning a local continuous flux level of 30 [10] counts per sample interval in the 1-9.5 [9.5-73] keV bandpass, we obtain a fluence of 175 [90] counts in the FWHM interval of the event in that bandpass. The spectral hardness ratio of the (1-9.5)/(9.5-73) keV fluences is 1.9 ± 0.3, where we assume a √N Poisson distribution uncertainty on the fluence. The continuous flux has a spectral hardness ratio averaged over the 100 s data set of 2.1, so the spectrum of the burst is consistent with the time averaged spectrum of the source. The FWHM, $\Delta\tau$, of this pulse is 24 ± 4 ms, corresponding to a light ravel time $c\Delta\tau$ = 7300 ± 1200 km.

### 3.2 Event 961217

The event occurred on 1996 December 17 at 22:12:18 UT, when Cyg X-1 was in the low state. The counting rate in a 1-13 keV bandpass is shown in Fig. 2, binned at 1.22 ms sample time. The event peaks at 17 counts in sample interval 77,044, where the local continuous flux level is 4 counts per sample interval. The Poisson probability of observing a sample interval with 17 or more counts in a data set with a mean of 4 counts per sample is $9 \times 10^{-7}$. There are 81,920 sample intervals in this data set, so the expectation value for the number of sample intervals with a counting rate this large or larger in this data set is 0.07.

Additional evidence supports the identification of this event as a burst of radiation form Cyg X-1. The event appears at a similar counting rate in all 10 phases of the data binned at 1.22ms. An increase in counting rate with similar morphology occurs at the same time in both the 1-5 keV bandpass and the 5-13 keV bandpass. The event shows a smooth rise to and fall from maximum when examined with higher time resolution (Fig. 3).

The FWHM of the pulse is 1.2 ± 0.6 ms, during which time its fluence exceeds the continuous flux by a factor of 3 in the 1-13 keV bandpass. The FWHM corresponds to a light travel time $c\Delta\tau$ = 370 ± 150 km. The (1-5)/(5-13) keV spectral hardness ratio of the fluences in the pulse is 1.4 ± 0.8. The spectral hardness ratio of the continuous flux in the same bandpasses, averaged over the 100 s data set, is 0.6. The small fluences in the pulse make a comparison difficult, but they do allow the spectral hardness ratio of the pulse to be consistent with the spectral hardness ratio of the source's continuous flux.

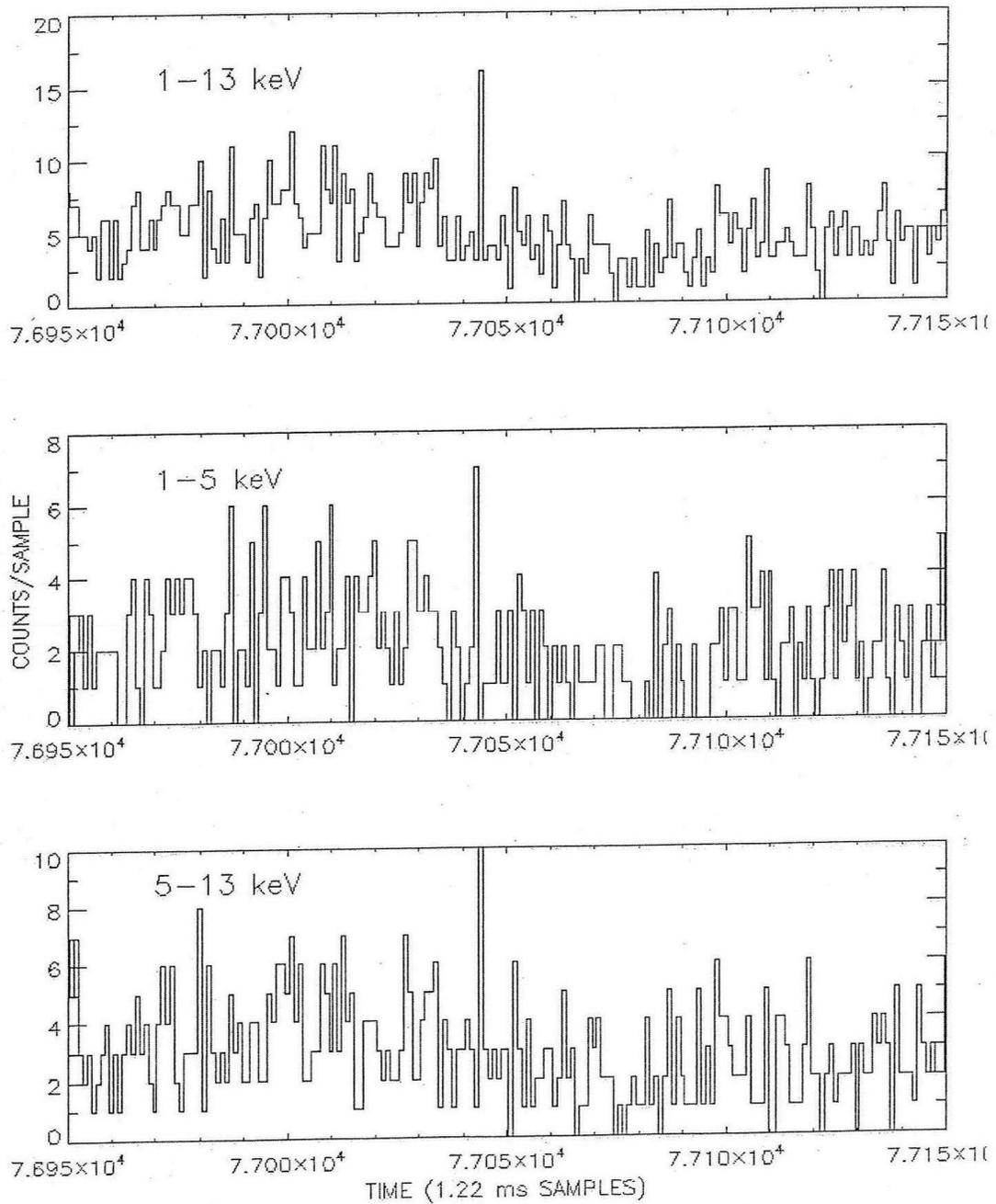

Fig. 2.— Event 961217. The PCA counting rate at 1.22 ms sample time. The event peaks in sample interval 77,044. *above*: 1-13 keV bandpass, phase 1; *middle*: 1-5 keV, phase 8; *below*: 5-13 keV, phase 1.

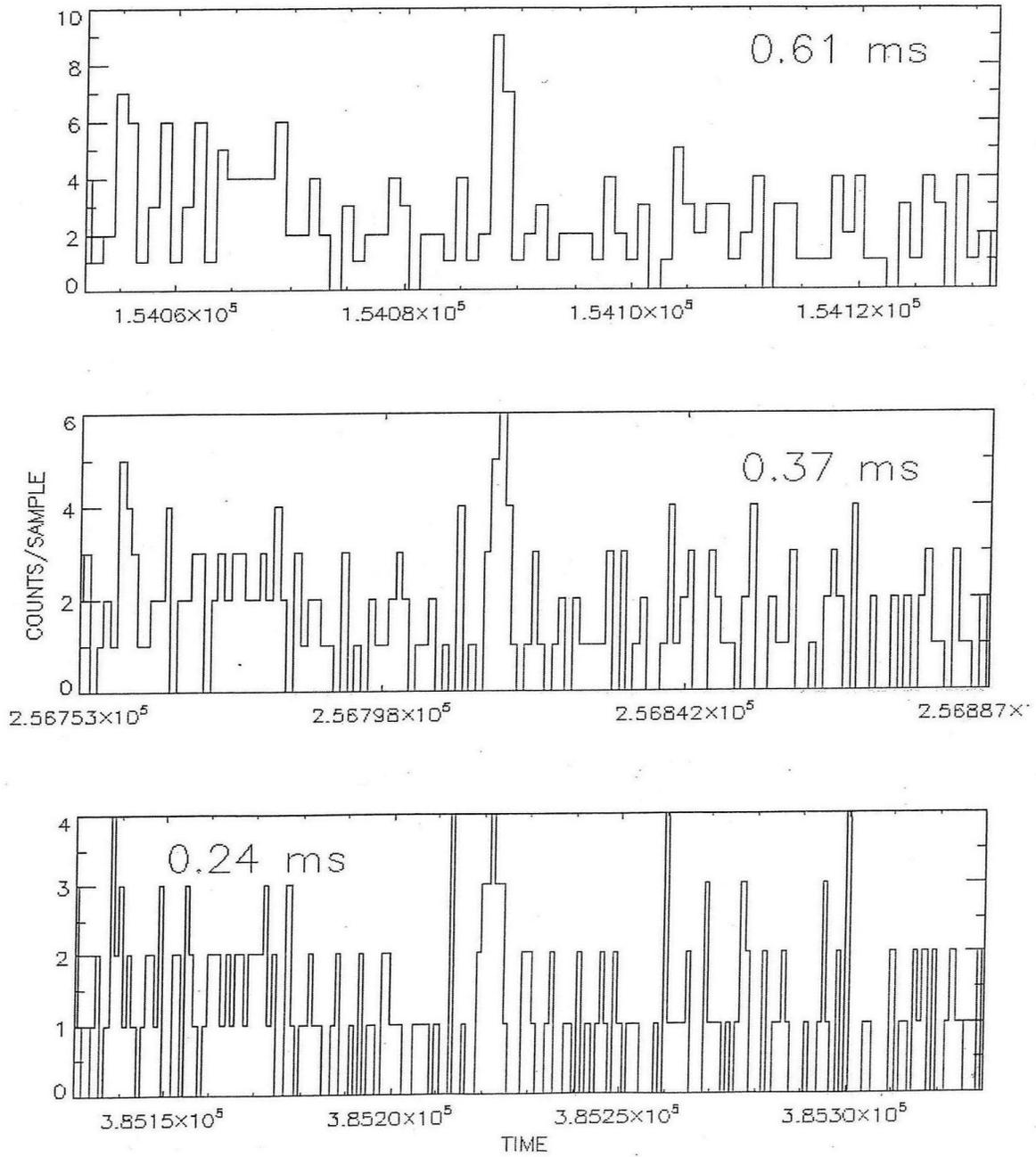

Fig. 3. Event 961217. The PCA counting rate in the 1-13 keV bandpass at (*above*) 0.61 ms sample time, phase 1; (*middle*): 0.37 ms sample time, phase 0; (*below*): 0.24 ms sample time, phase 1. The same 49 ms interval centered on the event is shown for each time resolution.

### 3.3 Events 990925A and B

The events occurred on 1999 September 25 at 16:05:48 UT, when the source was in the process of making a failed transition from the low state to the high state. The counting rate in a 2-15 keV bandpass is shown in Fig. 4, binned at 4.99 ms sample time. The event we label A peaks at 59 counts in sample interval 3271; event B peaks at 53 counts in sample interval 3280. The local continuous flux level is 20 counts per sample interval. The expectation value for the number of sample intervals with a counting rate this large or larger in this data set is 4 x $10^{-8}$ [2 x $10^{-5}$].

The morphology of the two events is similar at all 20 phases of the data with a 4.88 ms sample time. The two events appear similar in shape and occur at the same time in both the 2-6 keV bandpass and the 6-15 keV bandpass. Both events show the characteristic rise to and fall from maximum of a pulse when examined with faster time resolution. We identify both events as pulses.

The separation between events A and B is 44 ± 4 ms. The FWHM of A [B] is 24 [19] ± 4 ms, corresponding to a $c\Delta\tau$ = 7800 (5800) ± 1200 km. We note that the counting rate in A at 4.88 ms time resolution has a local maximum on the rising side of the pulse 2 sample intervals before the true maximum. The FWHM of both A and B is 10 ± 2 ms at 2.44 ms time resolution, corresponding to $c\Delta\tau$ = 3000 ± 1200 km if one exludes the precursor local maximum.

The fluence in both pulses in the 2-15 keV bandpass is equal to the continuous flux from the source over the duration of their FWHM. The (2-6)/(6-15) keV spectral hardness ratio of A [B] is 1.9 ± 0.3 [2.0 ± 0.4], significantly softer than the spectral hardness ratio of 1.2 derived from the continuous flux averaged over the 300 s data set.

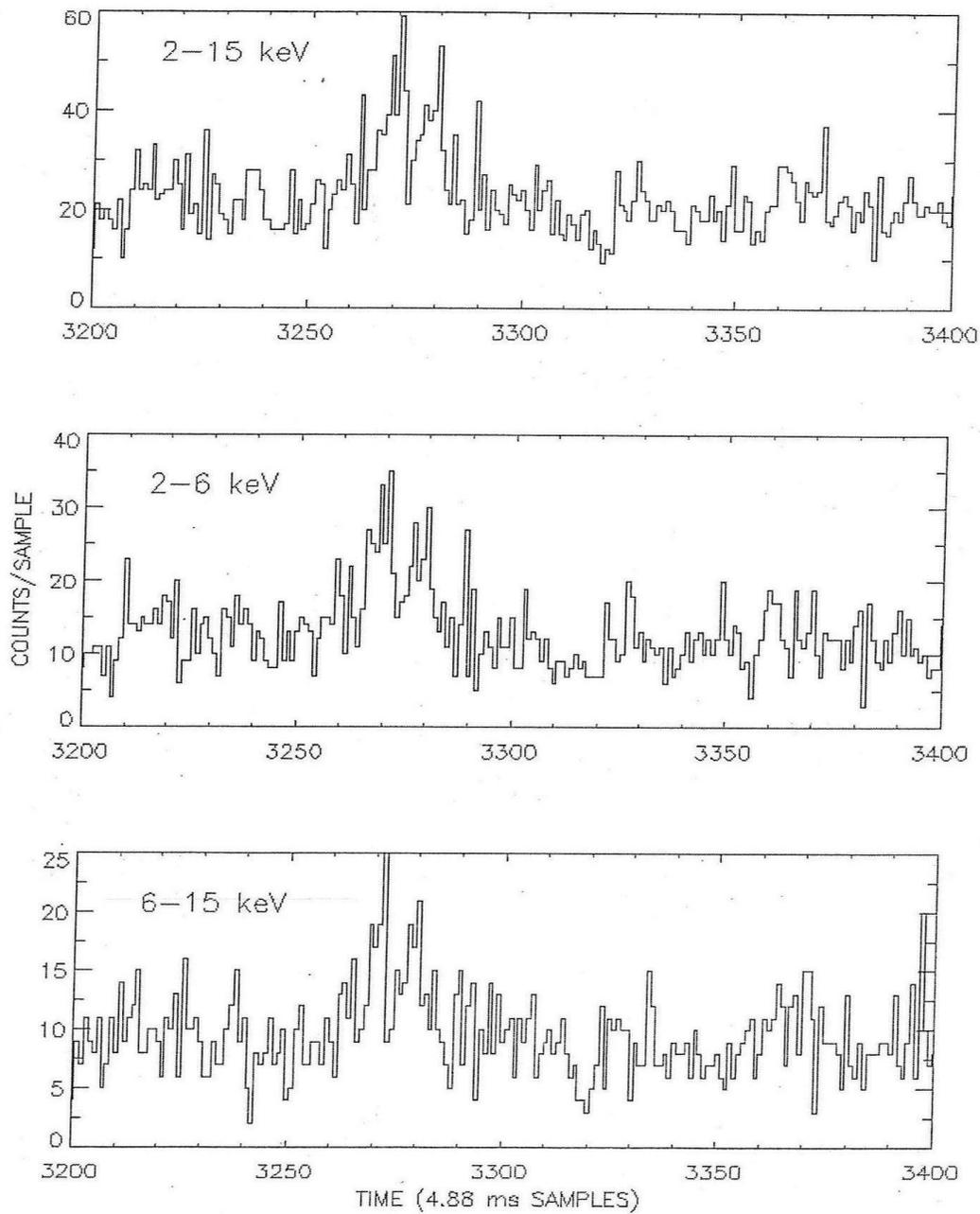

Fig. 4. Events 990925A and B. The PCA counting rate at 4.88 ms sample time. Event A peaks in sample interval 3271; event B, in sample interval 3280. *above*: 2-15 keV bandpass, phase 9; *middle*: 2-6 keV bandpass, phase 9; *below*: 6-15 keV bandpass, phase 5

### 3.4 Events 9909025C and D

These events also occurred on 1999 September 25, but at 18:02:57 UT, two hours after events 99025A and B. The counting rate in a 2-15 keV bandpass is shown in Fig. 5, binned at 4.88 ms sample time. The event we label C peaks at 51 counts in sample interval 33,075; event D peaks at 46 counts in sample interval 33,082. The local continuous flux level is 20 counts per sample interval. The expectation value for the number of sample intervals with a counting rate this large or larger in the 200 s data set is $1 \times 10^{-4}$ and 0.01, respectively. The morphology of the events is similar at all 20 phases of the data with 4.88 ms sample time. Similar events occur at the same time in both the 2-6 keV and 6-15 keV bandpasses. The separation between the two events is $34 \pm 4$ ms.

When examined with a sample time of 2.44 ms (Fig. 6), both C and D appear to be composed of two separate maxima in their counting rate. In the chronological order in which they occur, we denote the peaks as C1, C2, D1, and D2. The double pulse structure is visible when the data is examined at all sample times down to 0.244 ms, the time resolution of the observations. C1, C2, D1, and D2 all have the rise to and fall from maximum characteristic of a pulse at all sample times shorter than 2.44 ms. In the data with the fastest sample times, the separation between C1 and C2 is $6.8 \pm 0.5$ ms, and between D1 and D2, $4.4 \pm 0.5$ ms.

The FWHM of both C and D when examined with a sample time of 4.88 ms is $15 \pm 4$ ms, corresponding to $c\Delta\tau = 4400 \pm 1200$ km. The FWHM of C1, C2, D1, and D2 when examined with a sample time of 1.22 ms is 3.7, 2.4, 2.4, and 1.2 ms respectively, all $\pm 0.9$ ms. These FWHM correspond to a light travel time $c\Delta\tau$ ranging from $1100 \pm 270$ to $360 \pm 270$ km. The fluence in both C and D in the 2-15 keV bandpass equals the continuous flux from the source over the duration of their FWHM. The (2-6)/(6-15) keV spectral hardness ratio of C [D] is $0.6 \pm 0.2$ [$0.35 \pm 0.25$], significantly harder than the spectral hardness ratio of 1.2 derived from the continuous flux averaged over the 200 s data set containing the pulses.

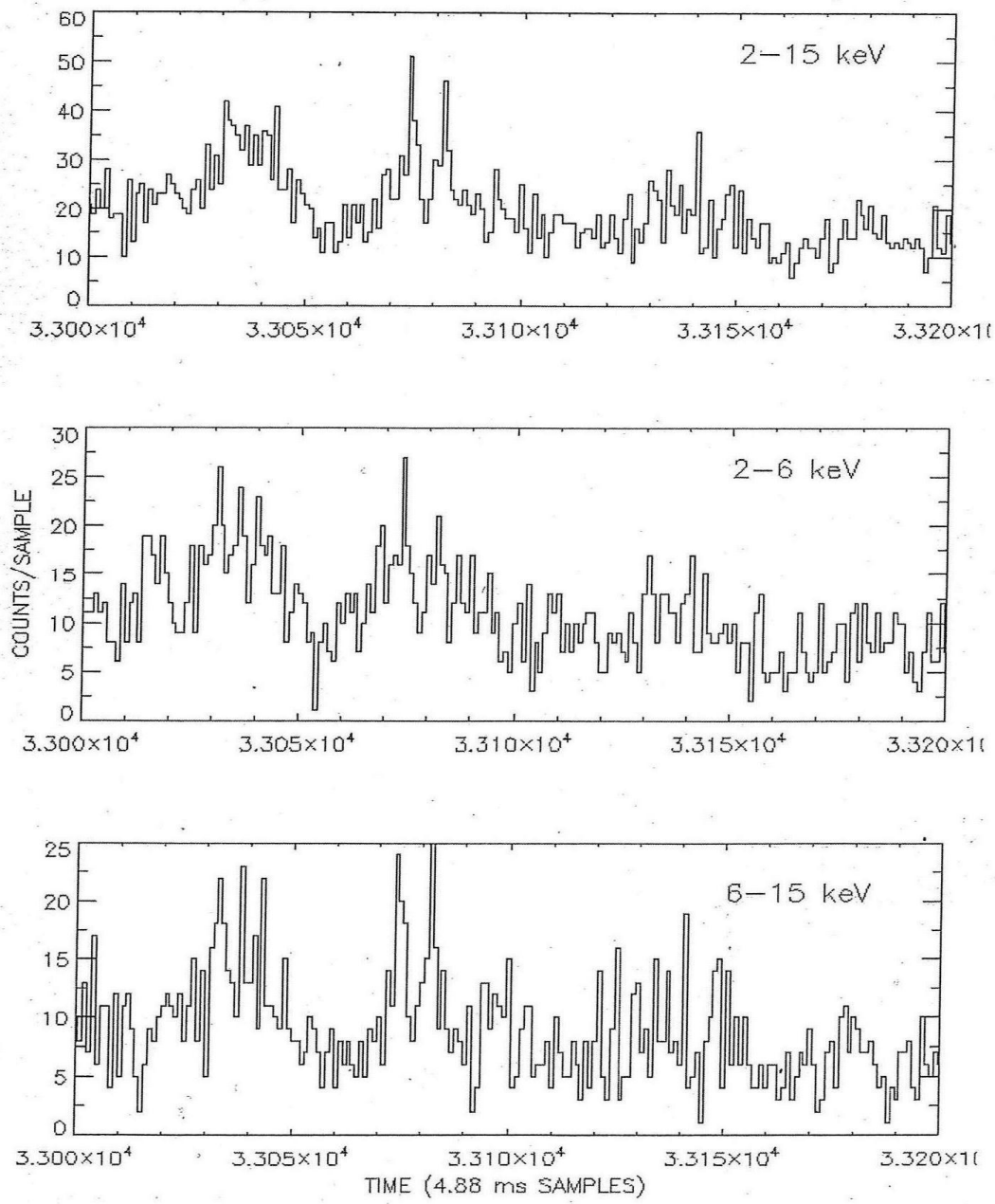

Fig. 5. Events 990925C and D. The PCA counting rate at 4.88ms sample time. Event C peaks in sample interval 33,075; event D, in sample interval 33,082. *above*: 2-15 keV bandpass; *middle*: 2-6 keV bandpass; *below*: 6-15 keV bandpass, all phase 5

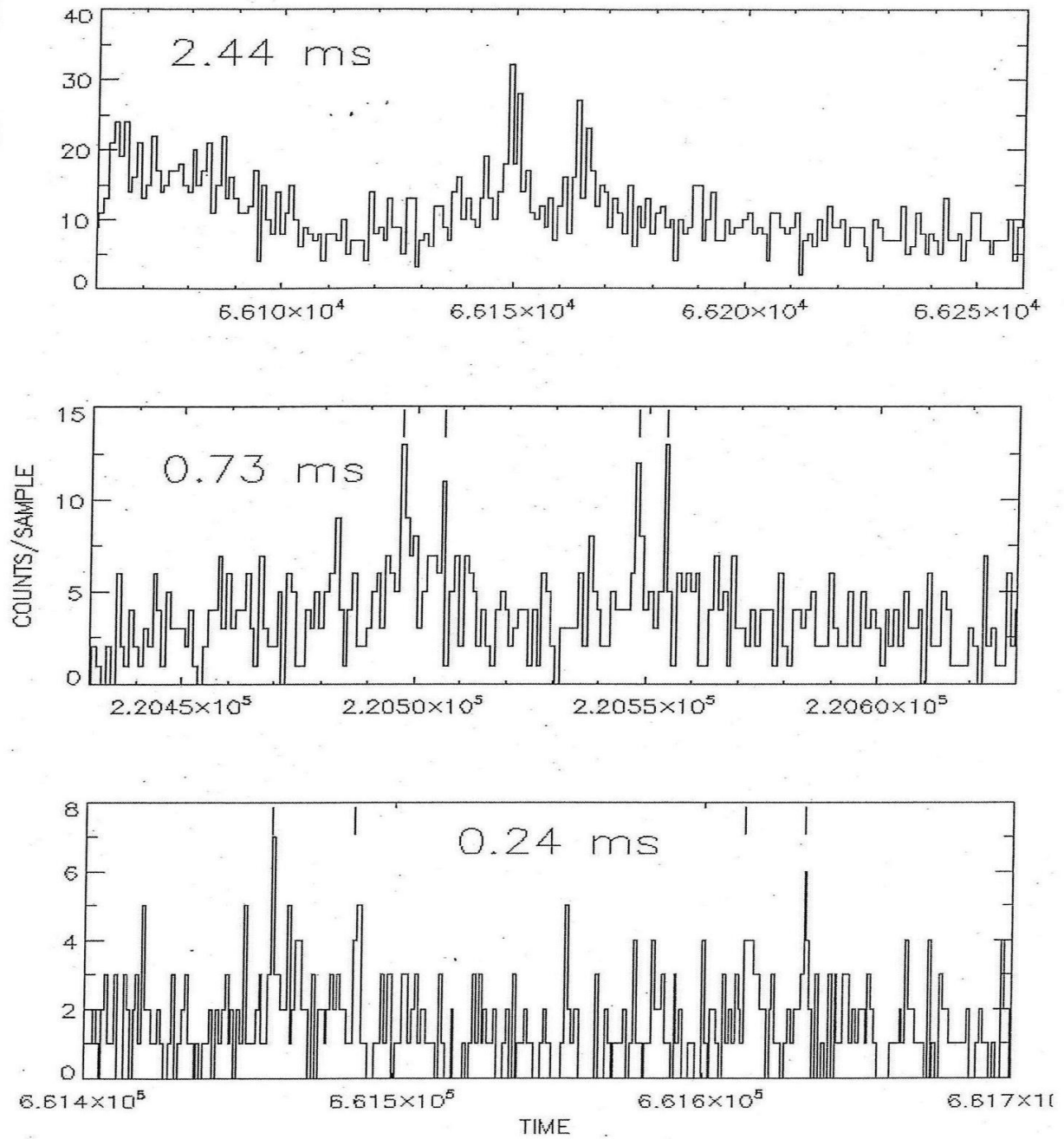

Fig. 6. Events 990925C and D. The PCA Counting rate in the 2-15 keV bandpass at (*above*) 2.44 ms sample time; (*middle*) 0.73 ms sample time; (*below*) 0.24 ms sample time (the time resolution of the observations), all phase 0. Each panel shows 200 sample intervals, so the linear separation of events C and D in a panel increases with increasing time resolution. "Sub-pulses" C1, C2, D1, and D2 are indicated by the vertical line above them.

## 3.5 Event 960811

The event occurred on 1996 August 11 at 07:22:55 UT, 2 days before Cyg X-1 underwent a transition from the high state to the low state. The counting rate in a 1-13 keV bandpass is shown in Fig. 7 binned at 2.44 ms sample time. The event peaks at 75 counts in sample interval 28,992; the local continuous flux level is 40 counts per sample interval. The expectation value for the number of sample intervals with a counting rate this large or larger in the 100 s data set is $8 \times 10^{-4}$. The morphology of the event is similar at all 20 phases of the data set, as are the events occurring at the same time in the 1-6 keV and 6-13 keV bandpasses.

The event exhibits a rise to and fall from maximum characteristic of a pulse when examined in the 1-13 keV bandpass with faster sample times. Its FWHM at 0.61 ms time resolution is $1.2 \pm 0.9$ ms, corresponding to a light travel time $c\Delta\tau = 360 \pm 270$ km. The (1-6)/(6-13) keV spectral hardness ratio of the pulse is $1.8 \pm 0.4$, consistent with the spectral hardness ratio of 2.3 derived for the continuous flux averaged over the 100 s data set.

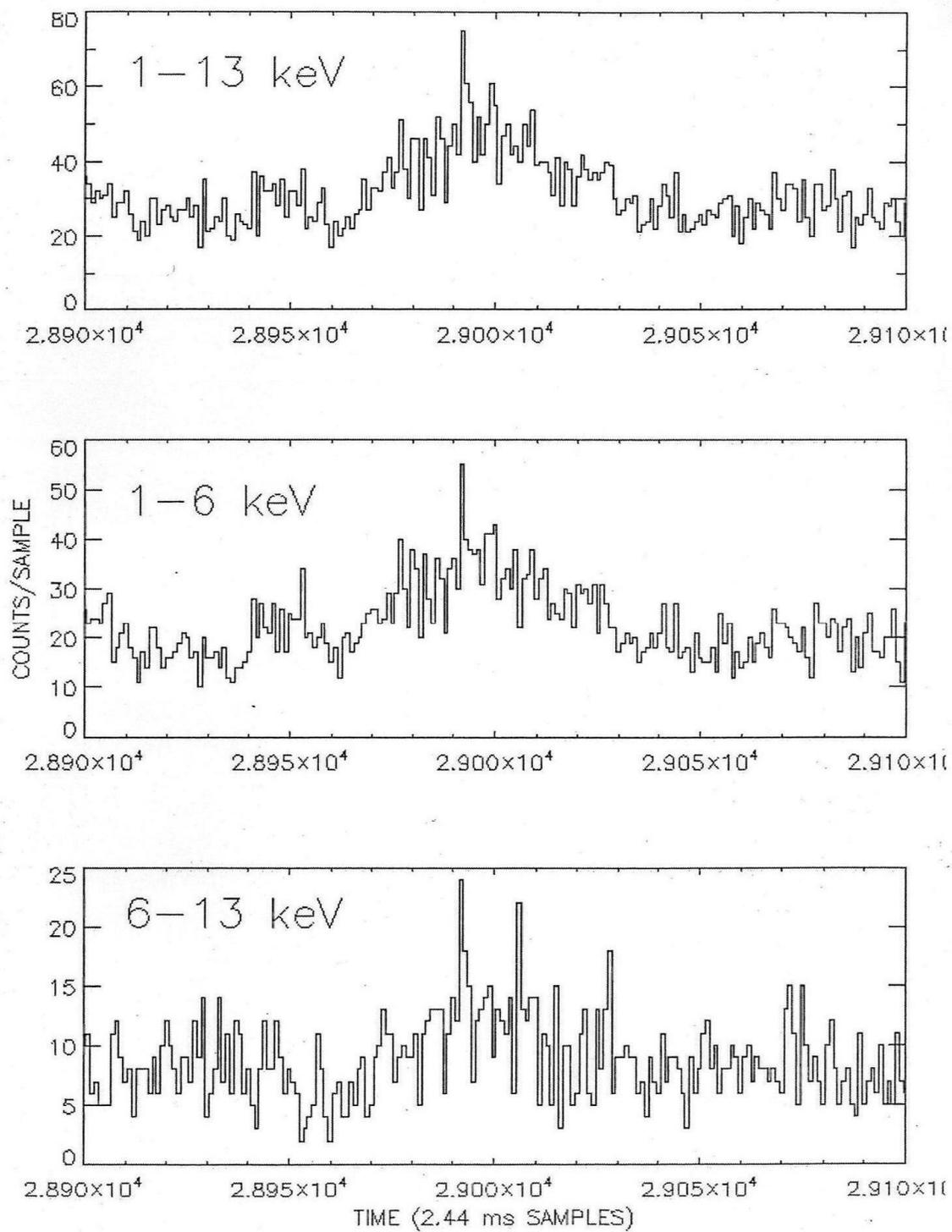

Fig. 7. Event 960811. The PCA counting rate at 2.44 ms sample time. The event peaks in sample interval 28,992. *above*: 1-13 keV bandpass, phase 5; *middle*: 1-6 keV bandpass. Phase 5; *below*: 6-13 keV bandpass, phase 10.

### 3.6. Event 960522

The event occurred on 1996 May 22 at 18:20:49 UT, 6 days after Cyg X-1 made a transition to the high state from a low state. The counting rate in a 1-13 keV bandpass is shown at 4.88 ms sample time in Fig. 8. The event peaks at 105 counts in sample interval 9381. With a local continuous flux level of 40 counts per sample interval, the expectation value for the number of sample intervals with this large a count or larger in the 100 s duration data set is $1.5 \times 10^{-19}$. The morphology of the event is similar at all 20 phases of the data set, as is the morphology of the event occurring at the same time in the 1-6 keV bandpass and the 6-13 keV bandpass.

The event exhibits a rise to and fall from maximum consistent with the characteristics of a pulse when examined at faster sample times. Its FWHM at 1.22 ms sample time is $7.3 \pm 0.9$ ms, corresponding to $c\Delta\tau = 2200 \pm 270$ km. The spectral hardness ratio of the fluences in the pulse in the (1-6)/(6-13) keV bandpasses is $5.1 \pm 0.8$, significantly softer than the spectral hardness ratio of 2.8 derived for the continuous flux averaged over the 100 s duration of the data set.

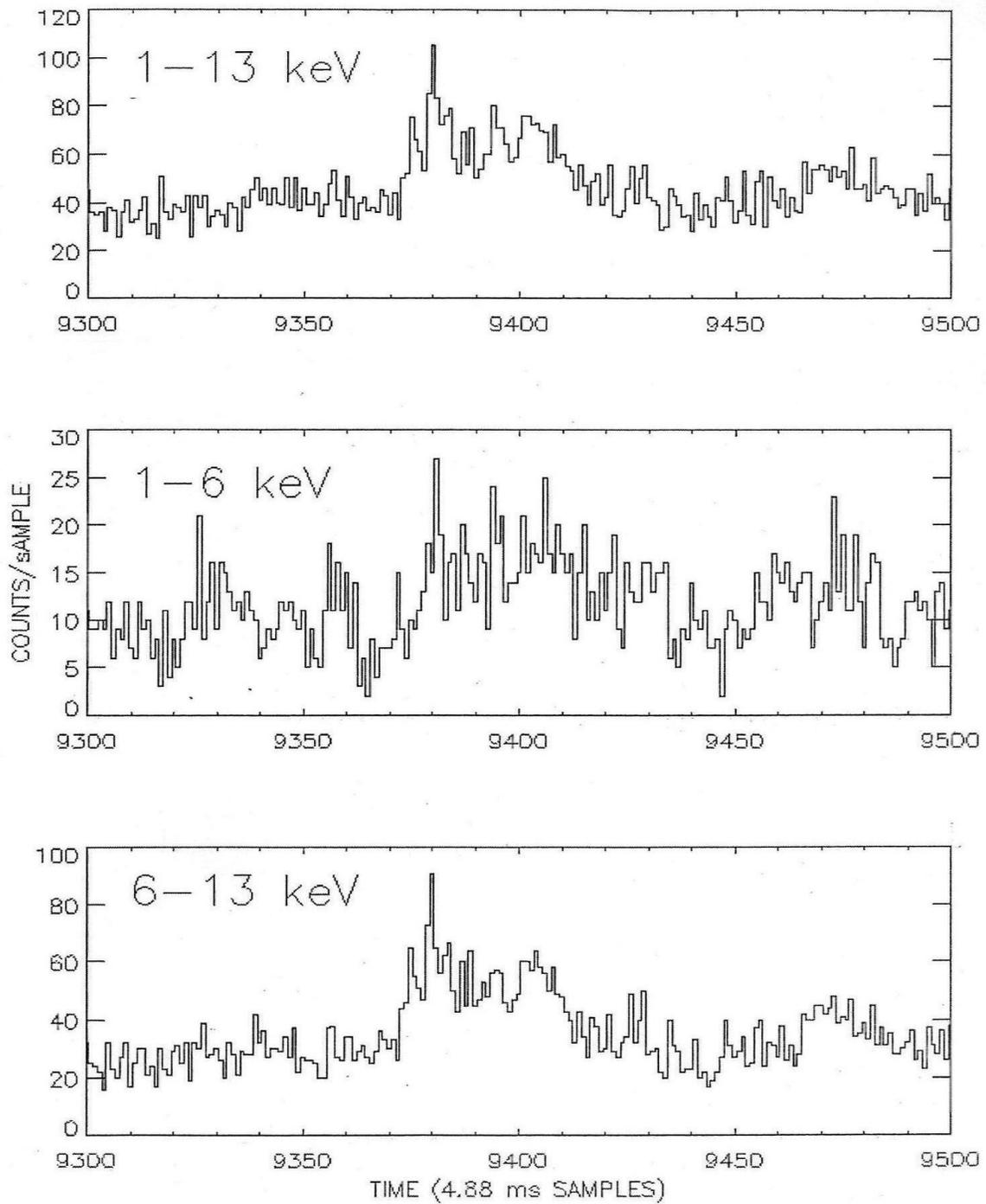

Fig. 8. Event 960522. The PCS counting rate at 4.88 ms sample time. The event peaks in sample interval 9381. *above*: 1-13 keV bandpass; *middle*: 1-6 keV bandpass; *below*: 6-13 keV bandpass, all phase 35.

## 4. DISCUSSION

The millisecond-timescale X-ray pulses in Table 1 are a small subset of similar events that occur in every RXTE observation of Cyg X-1 we examined. They were selected to illustrate the range of values in parameter space that the pulses possess. These events are statistically significant increases in the counting rate with FWHM in the 1 – 10 ms range. The events are evident in every energy bandpass we examined between 1 and 73 keV. They occur in the low luminosity state, the high luminosity state and during failed transitions between luminosity states. Their spectral hardness ratio is usually the same as that of the continuous flux, but it can also be significantly harder or softer for a specific pulse. The fluence of a pulse over the duration of its FWHM equals or exceeds the fluence of the continuous flux from the source. (This is a selection effect of the criteria used to identify a pulse. A population of pulses with fluence smaller than that of the continuous flux may also exist, but would not be identified with our criteria.) In short, RXTE observations resolve the historical discussion about the existence of ms-timescale pulses from Cyg X-1. X-ray pulses with ms-timescale FWHM are common in the radiation from the source.

The question arises as to why these pulses were not detected (or not identified as being more common) in previous studies. Several factors contributed to previous negative results. Not every factor occurred in every reported result, but at least one occurred in every case.

- As pointed out by Giles (1981), the lower counting rates associated with smaller area detectors used in rocket-borne experiments lowered the statistical significance (S/N ratio) of any bursts of radiation that did occur during the observation. This made pulses harder to detect in rocket-borne experiments. We analyzed data from the PCA on RXTE, in which Cyg X-1 produced a continuous counting rate of several detected photons per ms.

- Analyzing the data for pulses at a single sample time much shorter or much longer than the FWHM of ms-timescale pulses makes them appear to have a lower S/N than they do when analyzed with a sample time equal to their FWHM, as discussed in Sec. 2. Several previously reported results analyzed their data only at the sub-ms time resolution of their observations. Gierlinski and Zdziarski (2003), on the other hand, detected only 13 "fast flares" in over 600 hours of RXTE observations because they analyzed the data using a 125 ms sample time. These authors detected only the largest events that extended over > 100 ms timescales.

- Failing to analyze binned data at all possible phases introduces a bias against the detection of pulses, again as discussed in Sec. 2. This study is the only one that included the analysis of binned data at all phases of the binned data set.

- Individual pulses have markedly different characteristic timescales (cf. Table 1). Analysis of observational data in the frequency domain (i.e, using power spectrum analysis or auto-covariance analysis) may not detect significant power from these aperiodic events. We analyzed the data in the time domain, where individual events can be statistically significant.

Characteristic FWHM $\Delta\tau$ corresponding to light travel times $c\Delta\tau < 10^4$ km indicate that the ms pulses originate in the inner region of the accretion disk. Rothschild et al. (1974) show that flare patches near the innermost stable orbit around a black hole are a reasonable source of ms-timescale events.

Events 990925C and D are separated by 34 ± 4 ms. This is the timescale expected for the rotational period of material near the inner edge of an accretion disk around a several solar mass black hole. For a Schwarzschild (a = 0) black hole, this period is

$$P = 0.6 \, m \, (r/3)^{3/2} \text{ ms} \qquad [2]$$

where m and r have the meaning defined in Eq. [1]. The inner edge of an accretion disk around a Schwarzschild black hole is defined by the innermost stable orbit at r = 3. Identifying the pulse separation timescale as the rotation period of a flare patch around the black hole would mean that the two separate pulses come from the same flare patch. Several characteristics of the two events are consistent with their origin in the same flare patch viewed on two successive orbits around the black hole. Both pulses show a similar morphology with a double maximum structure (Fig. 6). The two "sub-pulses" (C1 and C2; D1 and D2) are separated by the same interval, 4.8 ± 1.7 ms. The measured FWHM and spectral hardness ratio of C and D are also consistent with both pulses having the same value. Despite these similarities, however, no definite conclusion can be made regarding the identification of these two pulse as events arising from the same flare patch viewed on two successive orbits.

X-ray pulses with a ms-timescale occur in every luminosity state of Cyg X-1. Because they originate in the inner accretion disk, they may provide a way of investigating the properties of the inner disk and of the geometry of space-time near the event horizon of a black hole.

Further spectroscopic, photometric, and polarimetric studies of ms pulses seem indicated for this purpose.